# Theory of Quantum Computing and Communication

A report from the NSF sponsored workshop held
January 17-18, 2002 in Elmsford, New York

# Introduction

For most of the history of computer science, researchers have considered information mostly from a binary point of view: bits are either true or false. Quantum mechanics tells us that these bits may lie in some "superposition" of true and false. Considering quantum mechanics fundamentally changes the way we must consider computation, communication and information in ways that we are only beginning to understand. We strongly recommend that the NSF Division of Computer-Communications Research (C-CR) develop a new initiative in "Theory of Quantum Computing and Communication" to understand these issues from two directions: how the theory communities can contribute to more effective techniques in quantum information processing and how the emerging availability of quantum information processing raises research questions central to the theory communities.

Computer science has played a critical role in the development of quantum information processing. Most notably, Shor's algorithm for efficient factoring on quantum computers has shown that quantum computation can give exponential speed-up on some natural problems. Computer scientists have also led the way to develop error-correcting protocols to handle decoherence problems in quantum computers.

In the sections that follow we give a detailed exposition of several important questions about quantum information that requires careful examination by computer science researchers. We divide these questions into four categories.
- Algorithms and Complexity: What is the power of quantum computation?
- Cryptography: How can quantum information lead to better cryptographic protocols?
- Information and Communication: How do entanglement and other properties of quantum bits allow us to improve communications and information storage?
- Implementations, Models and Applications: How can theoretical computer science help in the design and structure of quantum machines? What are the theoretical questions arising in the application of quantum mechanics to a wide variety of computer science applications?

We believe that C-CR should establish a new initiative in the "Theory of Quantum Computation and Communication" to promote and fund research studying these issues.

Based on a 1999 workshop held in Arlington, Virginia, the National Science Foundation released a report "Quantum Information Science: An Emerging Field of Interdisciplinary Research and Education in Science and Engineering". This report made the case for a broad interdisciplinary study in quantum information. The NSF QuBIC Program including support for work on quantum inspired computing from an interdisciplinary perspective may be considered to be an outcome of the 1999 report. Our recommendation for a theoretical computer and communications initiative in the quantum area can be considered as complementary to QuBIC.

A workshop sponsored by NSF C-CR and organized by the Center for Discrete Math and Theoretical Computer Science (DIMACS) was held at Elmsford, New York, January 17-

18, 2002 where we had several discussions that gave structure to this report. The workshop immediately followed the Fifth Annual Quantum Information Processing Workshop that was held January 14-17 at IBM Yorktown. This allowed us to draw leading researchers from the QIP workshop to help develop this report. A list of participants in the workshop appears in the appendix.

# Algorithms and Complexity

Quantum Algorithms such as Grover's algorithm for database search and Shor's algorithm for factoring have shown that quantum computers, once built, can greatly improve on traditional computing for some problems. The National Science Foundation should fund continual research into understanding the ultimate algorithmic power of quantum computing along three fronts: Developing new algorithms, understanding and developing new techniques used by quantum algorithms and optimizing and extending and optimizing current algorithms.

## Developing New Algorithms

Even if quantum algorithms cannot efficiently solve NP-complete problems, there remain several interesting questions, like factoring, that apparently do not have efficient solutions in classical computing but do not appear to be NP-complete either. These problems are good candidates for finding efficient quantum algorithms. These problems include

- Lattice problems and code problems: Problems in lattices have applications to cryptography and coding. Such problems include approximating the shortest vector in a lattice.
- Graph Isomorphism: Given two graphs describing a relational structure, are they equivalent up to relabeling? This problem has drawn attention as the most natural likely to have an efficiently quantum algorithm.
- Group Problems: Graph isomorphism is an example of a more general hidden subgroup problem. Do these and related group problems have good quantum solutions.
- Simulating Physics: How efficiently can quantum algorithms simulate quantum and classical physical systems?

In another direction we can look at smaller than exponential improvements in algorithms. If truly efficient quantum computers emerge than even with these smaller improvements we can handle problems of far larger size with quantum computers.

- Clever classical techniques have given faster than expected exponential algorithms for various NP-complete problems such as solving Boolean formula satisfiability on 3CNF formulas. Perhaps clever quantum techniques can get even better exponential bounds.
- Problems like maximum flow and linear programming have polynomial-time in the worst-case solutions but with relatively large exponents. Perhaps much more efficient quantum algorithms exist.

- Explore the power of QNC that represents what can be computed with qubits with a short life span. The greatest-common divisor problem might be a candidate for a problem not in NC but in QNC.

## Understanding quantum algorithmic techniques

We have seen several tools for developing quantum algorithms. We should understand the power and limitations of these techniques.

Adiabatic algorithms make a slow transition from one state to another to find an optimal solution. They have shown some promise in simulations and we should determine whether they could give great improvements in NP-complete problems in general.

Studying adiabatic and other heuristical algorithms may lead to quantum algorithms which may have bad worse case performance but nevertheless give efficient solutions in many practical settings.

The quantum Fourier transforms form the basis of nearly all the known quantum algorithms. Are there any other fundamentally different ways to apply quantum computing? What are the limitations of the quantum Fourier transforms?

Amplitude amplification allows us to reduce the error of a quantum computation dramatically with fewer than expected repetitions. Are there other similar techniques to reduce the error of quantum algorithms?

Taking a quantum random walk through a graph structure has surprisingly different properties than classical random walks. We should continue to study random walks and find applications for solving algorithmic problems.

In general, we should continue to discover and develop new techniques for quantum algorithms.

## Optimizing and Extending Current Algorithms

How can we improve the current quantum algorithms that already exist? These improvements can lead to implementations on early generation quantum computers.

- Can we reduce the number of entangled quantum bits needed for the algorithms?
- Can we adapt our algorithms to other architectures like the 1-clean qubit model that comes from NMR computing?
- Can we develop interesting algorithms for very small quantum computers? These algorithms could be implemented within a few years.
- How many qubits do we need for new problems?

## Complexity Theory

Quantum computing devices led to the development of computational complexity models based on models from these machines. How these classes relate to the classical classes remains an important area of study. Some of the more important classes:

- BQP – Problems efficiently computable on a quantum Turing machine. Is BQP in the polynomial-time hierarchy? Any interesting consequences of NP in BQP?
- QMA – Problems with efficiently verifiable quantum proofs. Can we put co-NP or graph non-isomorphism in this class?

- QIP – Quantum interactive proof systems. Is it equal to PSPACE, EXP or something in-between?
- QNC – Parallel small-time quantum class. Can we get P in here, i.e., parallelize classical computation by quantum means?

Besides the usual resources of time and memory, quantum machines have other resources that one might want to conserve to make an algorithm implementable. These resources include the amount of entanglement, the number of "clean" qubits, the number of measurements and the amount of nonuniform quantum advice.

We have seen considerable success in characterizing the number of queries made by quantum algorithms in the black-box model. Some questions like element distinctness still have large gaps in lower and upper bounds. Creating new lower bound techniques for this model could close the gaps in these problems.

Recent techniques have given time-space tradeoffs for some problems on classical machines. Can these or other techniques give interesting time-space tradeoffs for quantum machines?

Can we give an alternate characterization of quantum classes that one could use to analyze their relationship with classical classes? One can also give limitations on the power of quantum computation by looking at larger but more understandable classes.

How can the study of quantum complexity help in the analysis of classical complexity? We have already seen some examples of this:

- The log-rank conjecture for communication complexity holds in the quantum case.
- New methods for lower bounds for inner product in communication complexity.
- Cannot characterize classical property testing or decision-tree complexity by multivariate low-degree polynomials.

## Quantum Communication

Suppose Alice and Bob each have part of an input and would like to compute a function on the whole input. In some cases, the number of qubits needed for communication is far smaller than the number of classical bits for the same function. Many interesting issues remain for quantum communication complexity such as

- What is the power of using entangled qubits for communication?
- We can construct functions which provably give an exponential improvement over classical. Can we find natural examples of such functions?
- The complexity of set disjointness remains wide open.
- How does nonlocality affect communication? For example does quantum multiprover proof systems have different power if the noncommunicating provers have entangled qubits.
- Devise experiments to account for the detection loophole problem to disprove the hidden-variable theory.
- How does communication complexity relate with other complexity models like decision tree complexity and branching programs?

# Cryptography

Quantum cryptography has had the greatest implementation success in all of the area of quantum information processing. We have good protocols for secret key exchange. However, we need to understand the power and limitations for other aspects of quantum cryptography particularly in areas of authentication, anonymity, tracing and reliability/denial of service issues.

- Do there exist good quantum one-way and trap door functions?
- Is there a good notion of quantum zero-knowledge and what can be achieved by it?
- Can we build classical or quantum systems secure against quantum computers?
- Can we create cryptographic primitives and reductions from one primitive to another?
- What are the theoretical limits and practical techniques for quantum repeaters?
- Can we efficiently with quantum computation break classical private key systems like DES and AES? Grover's algorithm gives a quadratic improvement but can we do better?

# Information and Communication

How does quantum mechanics affect the way we study information and communication. We should consider theoretical limits as well as short-term goals that may have viable implementations.

## Error-Correcting Codes

The threshold theorem says that with assuming certain conditions on the error processes, a general quantum algorithm can be made fault-tolerant as long as the error is below some fixed threshold (currently estimated to be approximately $10^{-4}$). Some assumptions on the error process are necessary, but it is not clear whether we currently know the most general assumptions. Can these assumptions be widened to include more general classes of errors (such as those discussed by Alicki, Horodecki et al?).

Extend methods to create efficient quantum codes both for quantum information and for classical information. What are the theoretical limits of quantum error-correcting codes?

Can we create practical schemes for small codes? We should examine specific error models and continuous-variable codes.

Are there good algebraic and geometric codes for encoding quantum information into quantum bits? Almost all quantum error correcting codes discovered to date come from a class called additive codes, which are somewhat analogous to linear codes in classical coding theory. Are there non-additive codes, which improve over the best additive ones?

How can we exploit degenerate quantum codes, which do not have a classical analogue?

Create good 2-way entanglement purification protocols using error-correcting with feedback.

## Channel Capacity and Information Theory

How well can we send quantum and classical information through quantum channels?

How does prior entanglement help? Suppose we have entanglement without communication.

Examine the product-state capacity, $C_{Prod}$ developed by Holevo, Schumacher and Westmoreland. Is it additive? What if we have feedback?

Examine prior sender-receiver entanglement, $C_{EA}$. With and without feedback. With and without entangled inputs. Can we get a reverse Shannon theorem?

How well can we send quantum information? Examine quantum capacity with no feedback ($C_q$). Is this given by the coherent information? When is the coherent information additive; are the nonadditive examples we have discovered the rule or the exception? Suppose we have feedback ($C_{Q_F}$) or a classical side channel ($C_{q_2}$)? How does this relate to distillable entanglement?

What is the additivity of various capacity-like quantities, in particular, product state capacity? Are there new entropy inequalities?

## Information Theory

How can quantum information flow in a network or distributed environment? Will quantum Shannon theory be useful?

- Can we calculate the capacities of real channels? Estimations could also be useful.
- Can we do quantum state manipulation to achieve communication?
- Can we show that quantum channels give us little or no improvement on classical capacity? This would be disappointing but sill worth knowing.

It is possible to have a channel from A to B with 0 quantum capacity and a channel from B to A with 0 capacity yet combined we get some capacity. Perhaps this can be made useful in a network setting.

## Entanglement and Miscellaneous

Classify interchangeable resources such as channels, interactions and states.
Characterize LOCC protocols.
Examine the additivity problems for measuring entanglement and channel capacity.
Define, classify and relate various notions of quantum capacity and entanglement measures.

Develop practical schemes for entanglement transformation.

Give a unified systematic theory of entanglement, including multiparty entanglement.

Use ppt-preserving entanglement transformations to bound distillable entanglement.

# Implementations, Models and Applications

We need to find a synergy between abstract models of computation and proposed implementations. As physicists better understand the limitations of what we can implement, computer scientists can better devise models and algorithms to handle these limitations. Conversely, as computer scientists understand what resources are critical for quantum algorithms to work, physicists can better design implantations to address these issues.

Consider the Knill-Laflamme Milburn theory of linear optics with feedback. This is a continuous variable computation that is leading to implementation questions. This model uses bosonic computation that gives an interesting counterpart to Valiant's theoretical model that is equivalent to fermonic computation. Valiant's result shows that the fermonic computation gives no additional power over classical computation.

Adiabatic computation started with physical ideas for implementation and has also led to interesting theory.

The NMR mixed-state model has limited clean qubits. We need to understand how powerful this model is.

What is a model for quantum computation? We need to properly define the state space and computational primitives. This gives rise to questions about the relative computational power of these models. Different models might have polynomial speed-up/slow-down between them. For each model we have to understand fault tolerance, optimization of programs and other issues.

There are several models and implementations that need investigation such as

- Anyons, which have natural fault-tolerance properties.
- Briegel's model of clustered states with measurement.
- What is the power of measurement in general?
- Optical lattices.
- Linear optics with various restrictions.
- Quantum cellular automata
- Hamiltonian-based interactions. What is the power of various Hamiltonian primitives for building unitary/transformation gates?
- Continuous-variable computation.
- The quantum Clifford group. Operators in the Clifford group can be easily implemented fault-tolerantly though any sequence of Clifford group operators can be simulated classically.

How can quantum mechanics be harnessed for image and signal processing? Technology is coming available for using quantum effects and/or entangled photons for imaging/microscopy/tomography. These can use simultaneous multiparameter entanglement with lead dispersion cancellation and better resolution.

What else can be done with this technology? Will quantum image processing, using new algorithms, make better use of this technology? This needs a theoretical framework that we can then analyze.

We need quantum feedback and control to manipulate quantum states and give us robust quantum computing. Algorithmic cooling is one example of this that makes specific quantum bits from a larger number of "messy" bits. This has proven useful in current experiments.

In general, we should take primitives and needs from implementation, model them formally and study the power of these models.

# Workshop Participants

| | |
|---|---|
| Julia Abrahams | National Science Foundation |
| Dorit Aharonov | California Institute of Technology |
| Andris Ambainis | Institute for Advanced Studies |
| Howard Barnum | Los Alamos National Laboratory |
| Charles Bennett | IBM Yorktown |
| Harry Buhrman | Centrum voor Wiskunde en Informatica |
| Richard Cleve | University of Calgary |
| Yvo Desmedt | Florida State University |
| Lance Fortnow | NEC Research Institute |
| Daniel Gottesman | University of California at Berkeley |
| Mark Heiligman | National Security Agency |
| Christopher King | Northeastern University |
| Debbie Leung | IBM Yorktown |
| Michael Nielsen | University of Queensland |
| Eric Rains | AT&T Labs-Research |
| Vwani Roychowdhury | University of California at Los Angeles |
| Leonard Schulman | California Institute of Technology |
| Alexander Sergienko | Boston University |
| Peter Shor | AT&T Labs-Research |
| Robert Sloan | National Science Foundation |
| Umesh Vazirani | University of California at Berkeley |
| John Watrous | University of Calgary |
| Andreas Winter | University of Bristol |